%% file: github_measurement.tex
\newcommand{\PaperTitle}{Analysis of Commit Signing on GitHub}
\newcommand{\PaperNumber}{22}

\documentclass[10pt,sigconf,letterpaper,nonacm]{acmart}
\hypersetup{
  colorlinks=true,
  linkcolor=blue,
  citecolor=blue,
  urlcolor=blue
}

\usepackage{amsmath,amssymb,amsfonts}

\usepackage{graphicx}
\usepackage{xcolor}
\usepackage{tikz}
\usepackage{adjustbox}
\usepackage{float}
\usepackage{pdfpages}
\usepackage{placeins}

\usepackage{booktabs}
\usepackage{tabularx}
\usepackage{longtable}
\usepackage{array}
\usepackage{colortbl}
\usepackage{ragged2e}
\usepackage{makecell}
\usepackage{multirow}
\usepackage{multicol}
\usepackage{threeparttable}
\usepackage{siunitx}

\usepackage{listings}
\usepackage{fvextra}
\DefineVerbatimEnvironment{wrapverb}{Verbatim}{
  breaklines=true,
  breakanywhere=true,
  fontsize=\small
}
\usepackage{textcomp}
\usepackage{upquote}
\usepackage{xurl}
\usepackage[inline]{enumitem}
\usepackage{microtype}
\microtypecontext{spacing=nonfrench}
\usepackage[all]{nowidow}
\usepackage{framed}
\usepackage{dirtytalk}
\usepackage{csquotes}

\usepackage[ruled,vlined]{algorithm2e}
\usepackage{algorithmic}

\usepackage[most]{tcolorbox}

\newcommand{\PP}[1]{%
  \par\addvspace{\medskipamount}%
  \noindent\textbf{#1}%
}

\definecolor{vtAlignGreen}{RGB}{39,80,10}
\definecolor{vtAlignBg}{RGB}{234,243,222}
\definecolor{vtDivergeRed}{RGB}{121,31,31}
\definecolor{vtDivergeBg}{RGB}{252,235,235}
\definecolor{vtPartialOrange}{RGB}{99,56,6}
\definecolor{vtPartialBg}{RGB}{250,238,218}
\definecolor{vtUniqueBlue}{RGB}{12,68,124}
\definecolor{vtUniqueBg}{RGB}{230,241,251}
\definecolor{vtSectionGray}{RGB}{235,233,228}

\newcolumntype{P}[1]{>{\raggedright\arraybackslash}p{#1}}

\emergencystretch=\maxdimen
\definecolor{utkorange}{HTML}{FF8200}

\newcounter{takeawaycnt} 
\newenvironment{takeawaybox}{%
    \setlength{\topsep}{0pt}\setlength{\partopsep}{0pt}\setlength{\OuterFrameSep}{0pt}%
    \def\FrameCommand##1{\setlength{\fboxsep}{0pt}\colorbox{gray!10}{\hbox{{\color{utkorange}\vrule width 2pt}\hspace{3pt}\vbox{\kern 1.5pt ##1\kern 1.5pt}\hspace{2pt}}}}%
    \MakeFramed{\advance\hsize by -7pt\relax \FrameRestore \setlength{\parindent}{0pt}}%
}{\par\endMakeFramed}

\newenvironment{mytakeaway}
    {\begin{takeawaybox}\textbf{Summary \thetakeawaycnt:} \stepcounter{takeawaycnt}}
    {\end{takeawaybox}}

\definecolor{quote}{rgb}{0.99,0.99,0.99}
\definecolor{bar}{rgb}{0.5,0.5,0.5}

\newcommand{\longsay}[1]{%
  \def\FrameCommand{%
    \hspace{2pt}%
    {\color{bar}\vrule width 2pt}%
    {\color{quote}\vrule width 4pt}%
    \colorbox{quote}%
  }%
  \MakeFramed{\advance\hsize-\width\FrameRestore}%
  \begin{list}{}{%
    \setlength{\topsep}{0pt}
    \setlength{\leftmargin}{0pt}
    \setlength{\rightmargin}{0pt}
    \setlength{\partopsep}{0pt}
  }
  \item[]
   {#1}%
  \end{list}%
  \endMakeFramed%
}

\def\BibTeX{{\rm B\kern-.05em{\sc i\kern-.025em b}\kern-.08em
    T\kern-.1667em\lower.7ex\hbox{E}\kern-.125emX}}

\begin{document}

\title{\PaperTitle}

\author{Abubakar Sadiq Shittu}
\orcid{0000-0002-4699-3015}
\affiliation{%
  \institution{University of Tennessee}
  \city{Knoxville}
  \state{TN}
  \country{USA}%
}
\email{ashittu@vols.utk.edu}

\author{Parker Collier}
\authornote{Contributed to this work while at the University of Tennessee.}
\orcid{0009-0000-1901-6326}
\affiliation{%
  \institution{atsec information security}
  \city{Austin}
  \state{TX}
  \country{USA}%
}
\email{parker.collier@atsec.com}

\author{John Sadik}
\orcid{0009-0007-8988-5501}
\affiliation{%
  \institution{University of Tennessee}
  \city{Knoxville}
  \state{TN}
  \country{USA}%
}
\email{jsadik@vols.utk.edu}

\author{Farzin Gholamrezae}
\orcid{0009-0008-1225-0329}
\affiliation{%
  \institution{University of Tennessee}
  \city{Knoxville}
  \state{TN}
  \country{USA}%
}
\email{fgholamr@vols.utk.edu}

\author{Scott Ruoti}
\orcid{0000-0002-6917-4186}
\affiliation{%
  \institution{University of Tennessee}
  \city{Knoxville}
  \state{TN}
  \country{USA}%
}
\email{ruoti@utk.edu}

\input{sections/0.abstract.tex}

\maketitle

\input{sections/1.new_intro.tex}

\input{sections/2.bg.tex}
\input{sections/3.rw.tex}
\input{sections/4.data-collection.tex}

\input{sections/5.data-analysis.tex}

\input{sections/7.results.tex}

\input{sections/6.limitation.tex}
\input{sections/8.discussion.tex}

\bibliographystyle{ACM-Reference-Format}
\bibliography{references}

\appendix
\input{sections/ethics.tex}

\section{Open Science}
\label{sec:open_science}

We have made our code for enumerating accounts, collecting public keys and commit data, and running our analysis available at \href{https://anonymous.4open.science/r/commit-signing-measurement/}{\textcolor{blue}{\nolinkurl{https://anonymous.4open.science/r/commit-signing-measurement/}}}. Because our dataset links GitHub profiles, keys, repositories, and commit metadata (including names and email addresses), we do not post the raw records as an open public download. Under GitHub's Acceptable Use Policies~\cite{github_aup}, researchers must secure collected personal information, prevent profiling or unsolicited contact, and honor removal requests, which cannot be enforced on an open public link. To support reproducibility while meeting these obligations, we store the full dataset on a restricted access institutional server and share it with researchers upon request. Requesting researchers must agree to use the data only for scientific research, keep it secure, honor removal requests, and not contact developers without ethical oversight.

\section{Generative AI}
\label{sec:Generative_AI}

We acknowledge the use of Grammarly AI solely for grammar, spelling, and light copy editing of our writing~\cite{grammarly2026}. No scientific content, analysis, or original text was generated by AI tools. We reviewed all edits suggested by Grammarly AI and take full responsibility for the content of this paper.

\clearpage
\input{sections/appendix_result.tex}

\end{document}

%% file: sections/0.abstract.tex
\begin{abstract}
Since 1999, the security community has known that managing cryptographic keys is hard for users of cryptographic systems.  Git lets software developers use private keys they control to digitally sign the commits that record their code changes to a repository. This practice offers a rare opportunity to study how a large community manages cryptographic keys in everyday work. In this paper, we study end-to-end commit signing, using GitHub accounts as our sampling unit. Our $2\%$ stratified sample of the GitHub population comprises $2,737,649$ accounts, $71,694$ of which have made at least one commit. For these accounts, we analyze $16,112,439$ commits across $874,198$ repositories and collect $208,546$ public keys uploaded to the sampled accounts. We examine developers' signing key usage patterns and compare our findings with those of prior commit signing studies, most of which used repository-based samples and paid less attention to key management. We then outline a path toward more usable key management for developers.
\end{abstract}

%% file: sections/1.new_intro.tex
\section{Introduction}
\label{sec:intro}

Public key cryptography is only as secure as the way users of cryptographic systems manage their cryptographic keys~\cite{diffie2022new,barker2018recommendation}. Yet ever since the seminal usability evaluation of PGP, it has been known that users struggle with key management, even when software follows conventional interface design standards~\cite{271563}. Nearly three decades later, effective key management remains a major bottleneck~\cite{10.1145/3830454.3846525}.

As a step toward addressing this multi-decade grand challenge, we believe it is important to examine communities that already manage cryptographic keys in practice and see if they are getting key management right. Software developers offer an ideal population for this investigation because version control systems like Git enable developers to cryptographically sign source code using keys they control, a practice we call \emph{end-to-end (E2E) commit signing}. Accordingly, we ask: \textbf{RQ1: What key management patterns can we observe in public GitHub artifacts?}

To answer this question, we began this research in 2023 by analyzing key management across a 2\% stratified sample of the GitHub population, comprising 2,737,649 accounts. From this sample, we constructed two datasets. First, we filtered for accounts with at least one public commit (71,694 accounts). For these accounts, we collected a list of 874,198 public repositories to which they contributed and identified 16,112,439 commits authored by them in those repositories. Second, we collected all public keys registered across the full sample of 2.7M accounts, identifying 208,546 keys owned by 122,050 accounts. Analyzing these datasets, key findings include:

\begin{itemize}
    \item Most signed commits matched public keys uploaded to GitHub, with exact fingerprint matches covering 82.15\% of E2E signed commits.
    \item About 60\% of accounts used only one signing key. Among accounts with multiple keys, 54.10\% switched keys and never returned to an earlier one, while 32.99\% reused older keys, sometimes within the same repository.
    \item Among accounts with multiple keys across multiple repositories, reusing keys across repositories and using multiple keys within a single repository were both common.
    \item Weak keys and deprecated algorithms were rare overall, but some persisted in recent signatures. GPG signatures using DSA, SHA-1, or keys below 2048 bit RSA equivalent strength were still used to sign commits as recently as 2025 and 2026.
\end{itemize}

During our data collection and analysis, several other papers were published that explored Git commit signing behavior~\cite{holtgrave2025attributing,sharma2025commit,zhang2025s}. While these papers did not focus on key management patterns (RQ1), they do raise interesting questions about commit signing behaviors. However, as these papers all explored commit signing from a repository centric view, it was unclear whether these results generalized across GitHub, or whether they were specific to the types of repositories examined. To answer this question of generalization, we leverage our collected data. As our data is a user centric view of commit signing, we can establish the generalizability of findings by triangulating our user centric data against this related work's repository centric data. This is particularly important because Olszewski et al.~\cite{309718,10.1145/3719027.3765573} emphasize that testing findings under new conditions helps establish the scope of those findings. As such, we ask: \textbf{RQ2: What are the generalizable commit signing behaviors across GitHub?}

We found that while low overall E2E signing adoption does generalize, relying solely on repository level views led to two notable oversights in prior work. First, contrary to prior assumptions, GitHub does not digitally sign all web interface commits; why some web commits are signed while others remain unsigned is undocumented and requires remediation. Second, developers are far more varied in their signing activity than repository centric studies suggest, exhibiting substantially higher adoption of newer methods like SSH signing, frequently switching signing behavior across repositories, and often stopping signing for a while before starting again. We provide a detailed comparison in Table~\ref{tab:rq2-comparison} in Appendix~\ref{others}.

\PP{Significance.} These findings highlight limitations in the current commit signing model, and we propose actionable improvements for platforms and commit tools in \S\ref{sec:discussion}. Beyond these fixes, usable key management requires understanding why developers make the choices they do. Our data does not explain why they stop signing, what problems they run into, or how they manage their keys over time. Closing these gaps will require hearing directly from developers who sign commits. Future qualitative studies interviewing these developers will help uncover what enables contributors to keep signing and what causes them to lapse, informing the design of better tools.  Looking further ahead, researchers should compare these findings and improved tool evaluations with evidence from other cryptographic systems to identify generalizable principles for any system where users manage their own cryptographic keys.

In summary, we make the following contributions:
\begin{enumerate}
    \item We provide the first account sampled analysis of cryptographic key management and usage patterns for commit signing on GitHub.
    \item We test the broad generalizability of prior repository centric measurements of commit signing.
    \item We offer recommendations for the research community, commit tools, and hosting platforms based on our findings.
    \item We share our source code and make our dataset available to researchers upon request to support replication and future research on commit signing.
\end{enumerate}

%% file: sections/2.bg.tex
\section{Background}
\label{sec:threat}

In this section, we define E2E commit signing workflow from a developer's workstation, detailing its operational scope, security guarantees, and underlying threat model.

\PP{Workflow.} When a developer edits files in a local clone of the repository, those changes initially reside unrecorded in the working directory. After staging the changes, the developer asks Git to record them as a commit. Git creates a commit object that references the staged snapshot by its root tree hash, includes the hashes of any parent commits, captures author and committer metadata, and attaches a commit message~\cite{blischak2016quick,yelgundhalli2025rethinking}. Because Git records the author name and email address from user-controlled settings, these fields are inherently self-reported~\cite{amreen2020alfaa}. Assuming the private key was generated securely, Git then signs the commit locally using the private key. This signature covers the entire commit payload including the tree hash, parent hashes, metadata, and message meaning that changing any of these fields causes verification of the original signature to fail~\cite{vaidya2019commit,yelgundhalli2025rethinking}. Finally, the developer pushes the signed commit to a remote repository such as GitHub. The hosting service authenticates the pusher, checks their repository permissions, and cryptographically verifies the signature against the developer's registered public key.

\PP{Threat Model.} We take source changes intentionally produced by trusted developers as our reference; assessing their correctness or freedom from vulnerabilities is outside our scope. We consider an attacker who seeks to introduce unauthorized source code changes into a project’s repository. They get write access by stealing a developer's access credentials or by bypassing authentication, as in recent GitHub supply chain compromises~\cite{cisa2025tjactions,unit42tjactions}. With write access, they can push any code with any author and committer name, because Git does not verify these fields~\cite{progit_signing_your_work,holtgrave2025attributing}. We treat the theft or compromise of a developer's signing key as an out-of-scope threat; we assume the attacker cannot steal such a key, make a developer's machine sign attacker-chosen content, break the verifier, or break the cryptography. We do not treat GitHub as malicious.

Maintainers keep a list of which keys belong to which trusted developers. A verifier checks every commit against this list and rejects unsigned commits, invalid signatures, and signatures from keys not on the list. Changing the list needs approval that a stolen access credential cannot give. This is necessary because a key found on an account is not trustworthy after a takeover: an attacker can upload their own key to a hijacked account~\cite{holtgrave2025attributing,prestonwerner2012publickey}. Git also reports whether a signature is valid separately from whether its key is trusted~\cite{gitdocs_prettyformats}.

Under these assumptions, repository access alone is not enough to create a commit with a valid signature from an authorized developer's key. The signature ties the commit's content and metadata to that key, so it supports attribution in a way that author and committer fields do not~\cite{progit_signing_your_work,slsa12source}.

GitHub signs commits made in its web interface with its own key~\cite{github_commit_signature_verification}. An attacker who takes over an account can use the web interface to get a GitHub-signed commit without the developer's key. This signature only shows that GitHub signed the commit. It is not the developer-key check we assume, so a generic ``Verified'' badge, or a policy that accepts any verified signature, is not enough~\cite{afzali2018legitimate,courtes2022guix,holtgrave2025attributing}.

These guarantees only hold if verification covers every commit the workflow accepts. They do not cover malicious authorized developers~\cite{wu2021hypocrite,james2024xz}, compromised signing machines, or attackers who can change the key list or skip verification. Signing also does not show that code is safe, and it does not stop rollback or other repository-metadata attacks~\cite{torres2016omitting}. It complements access control and code review.

%% file: sections/3.rw.tex
\section{Related Work}
\label{sec:rw}
%In this section, we review prior measurements on commit signing and public keys.

\PP{Commit Signing.}
Holtgrave et al.~\cite{holtgrave2025attributing} analyzed $25,889,629$ commits across $50,328$ dependent selected critical projects, separating user signatures from GitHub's \texttt{web flow}, and found that despite endorsement across $55$ security advice documents, $95.4\%$ of committers had never signed a commit, only $2.0\%$ signed all of theirs, and $72.1\%$ of projects had no signed commits. Adding a longitudinal view, Sharma et al.~\cite{sharma2025commit} examined $869,437$ commits across $60$ popular repositories in security, databases, web development, and machine learning with $15$ repositories in each category. Excluding bots and \texttt{web flow}, only $9.65\%$ of the $399,520$ human commits were verified, ranging from $28.35\%$ in security down to $2.55\%$ in machine learning, which rose after GitHub introduced verification badges and vigilant mode before declining in $2023$. They also compared four Git clients on signing and verification support and found that GitHub attributes commits to accounts via unverified email addresses. Scaling to the broader open source ecosystem, Mockus~\cite{mockus2026claimed} built a commit signature dataset for World of Code~\cite{ma2021world}, scanning $5,866,595,698$ commits to construct a per commit signature map, a key to author graph, and per identity trust tiers. Although $17.59\%$ of commits carried a signature, almost exclusively PGP at $98.96\%$, a key fanout gate separating shared keys from person keys showed that GitHub's \texttt{web flow} key alone links to over $30$~million authors, leaving only $0.74\%$ of canonical authors with a person attested key. Because signers skew toward clean developer identities, Mockus treats signatures as a precision anchor for identity resolution rather than a coverage metric.

Examining developer awareness and automated detection of email spoofing impersonation, Zhang et al.~\cite{zhang2025s,zhang2026understanding} combined repository mining, user studies, and authorship attribution. Across $12,577,332$ commits in $9,810$ top starred repositories where $56\%$ were unverified, metadata only cross checks failed to separate impersonation from benign merges and multi email workflows. Interviews with $17$ multi email contributors revealed prior unawareness of the threat, where mean concern rose from $3.9$ to $7.3$ out of $10$ after a live demonstration, and cited key management overhead as a signing barrier. To overcome missing metadata ground truth, their extended study~\cite{zhang2026understanding} reframed detection as repository specific authorship attribution using verified commits as cryptographic labels across $583,194$ commits spanning $2,156$ authors and $370$ repositories with at least three authors having five or more verified commits. CodeT5+ with commit metadata outperformed stylometry and few shot LLM baselines achieving $0.968$ mean per repository accuracy with minimal degradation under targeted style imitation attacks, providing a probabilistic triage signal that complements cryptographic signing.

\PP{Public Key Quality.}
Early audits of SSH keys uncovered widespread weak key generation. Cartwright-Cox~\cite{cartwrightcox2015auditing} profiled ${\sim}1.23$~million SSH authentication records via the Events API between December 2014 and January 2015, finding $256$-bit RSA entries and keys affected by the Debian OpenSSL randomness flaw. Shortly after, Barbulescu et al.~\cite{barbulescu2016rsa} crawled ${\sim}3$~million GitHub accounts between December 2015 and January 2016, of which $26\%$ held SSH-RSA keys, finding one factorable $512$-bit key alongside separate HTTPS X.509 factorization rates. Broadening these factorization checks, Cryptosense~\cite{cryptosense2015batchgcd} screened API-collected GitHub SSH keys for shared and small RSA factors from an unspecified corpus, fully factoring $12$ moduli of $2,048$~bits. Across $346$~million distinct X.509, SSH, and PGP keys, including $4.7$~million GitHub SSH, $1.2$~million GitLab SSH, and ${\sim}8,000$ GitHub PGP keys, Amiet and Romailler~\cite{amiet2018reaping} found shared RSA factors, ROCA-affected keys on both platforms, and cross-format reuse. More recently, B{\"a}umer et al.~\cite{10.1145/3719027.3765079} deduplicated public SSH client keys in June 2023 and January 2025 API snapshots from GitHub, GitLab, and Launchpad, counting $14.56$ and $17.06$~million distinct keys, with $9.20$ and $10.50$~million on GitHub and $5.78$ and $6.96$~million on GitLab, while preserving account and platform links. Evaluating algorithms, defects, and $24$ SSH clients, they showed GitHub's Ed25519 share rose from $37.46\%$ to $48.24\%$, while the January 2025 scan still found $98$ broken short keys, $139$ keys with weak randomness, and $149$ with common or small factors.

\PP{Research Gap.} The literature has either measured commit signing in popular repositories, where both individual developers' habits and repository specific requirements likely shape signing behavior, or examined key quality and algorithms without investigating how developers use their keys. To address this gap, we examine patterns of key use for commit signing across GitHub accounts.

%% file: sections/4.data-collection.tex
\section{Data Collection}
\label{sec:methodology}

%In this section, we describe how queried GitHub API endpoints to collect user accounts and their associated public keys. From these accounts, we identified those that had authored and pushed code changes. We then retrieved the repositories they contributed to, along with their individual commits within those repositories.

In this section, we describe how we used GitHub's APIs to sample accounts and collect their profile information and public keys. We also describe how we collected details about their commits, including any signatures, from the public repositories GitHub listed for them.

\subsection{Account Discovery}
\label{subsec:collection}

We constructed our dataset via GitHub's REST API \texttt{GET /users?since=\{id\}} endpoint~\cite{githubusersapi}, listing accounts in their order of creation to capture platform-wide adoption patterns. We retrieved profile attributes via the GraphQL API and all publicly available authentication and signing keys via dedicated REST endpoints~\cite{githubgraphqlapi,githubrestapi}. We collected these key categories separately because authentication keys support access to GitHub, whereas signing keys can be used to sign commits.

To scale collection, we pooled 10 Personal Access Tokens across researchers, using a dynamic rotation algorithm that monitors remaining quota and query complexity costs to maintain continuous, policy compliant throughput~\cite{githubgraphqllimits,githubbestpractices,githubratelimits}. GitHub's hard limit of 5,000 requests per hour per token~\cite{githubratelimits} makes full enumeration of all GitHub users, approximately 150 million accounts~\cite{githubabout}, infeasible (about 34 years at observed rates). This constraint likely explains why prior work restricted scope to specific repositories or public keys~\cite{zhang2025s,10.1145/3719027.3765079,holtgrave2025attributing,sharma2025commit}. GitHub declined a permanent rate-limit increase, so collection continued under the available request quotas. We therefore adopted stratified sampling aligned with established large-scale measurement practice~\cite{baltes2022sampling}.

\subsection{Account Sampling}

\PP{ID Space and Ceiling Estimation.}
GitHub’s account-listing endpoint returns accounts in sequential signup order, and identifiers are not recycled upon account deletion or administrative removal~\cite{githubusersapi}. The \texttt{GET /users?since=\{id\}} endpoint returns up to 100 users starting immediately after the specified identifier. To establish an upper sampling ceiling, we performed a manual threshold search by issuing queries at increasingly large values until responses became empty. Specifically, queries at \texttt{since}=\texttt{232M} and higher consistently returned empty responses, whereas queries near \texttt{230M} returned valid user objects, bracketing the boundary to the interval \texttt{[230M, 232M]}. We established approximately 230M as our working upper bound for this collection. This reflects allocated identifier space rather than active users; GitHub publicly reports approximately 150 million active accounts~\cite{githubabout}.

\begin{table}
  \caption{Collected data}
  \label{tab:collected-data}
  \centering
  \scriptsize
  \setlength{\tabcolsep}{3pt}
  \begin{tabularx}{\linewidth}{l l >{\raggedright\arraybackslash}X}
  \toprule
  Dataset & Field group & Fields \\
  \midrule
  \multirow{4}{*}{Accounts}
    & Identifiers        & \texttt{id}, \texttt{login}, \texttt{graphql\_id} \\
    & Profile            & \texttt{name}, \texttt{email}, \texttt{location}, \texttt{bio}, \texttt{company}, \texttt{website\_url}, \texttt{twitter\_username} \\
    & Account metadata   & \texttt{user\_type}, \texttt{site\_admin}, \texttt{createdAt}, \texttt{updatedAt} \\
    & Activity/graph     & \texttt{followers}, \texttt{following}, \texttt{repo\_count}, \texttt{repo\_contributed\_count}, \texttt{issue\_count}, \texttt{org\_count} \\
  \midrule
  \multirow{5}{*}{Public keys}
    & Identifiers        & \texttt{id}, \texttt{user\_id} \\
    & Key material       & \texttt{key}, \texttt{raw\_key}, \texttt{fingerprint}, \texttt{subkeys}, \texttt{key\_type}, \texttt{title} \\
    & Key lifecycle      & \texttt{createdAt}, \texttt{expiresAt}, \texttt{revoked}, \texttt{verified}, \texttt{read\_only}, \texttt{source} \\
    & Capabilities       & \texttt{can\_sign}, \texttt{can\_encrypt\_comms}, \texttt{can\_encrypt\_storage}, \texttt{can\_certify} \\
    & Associated identities & \texttt{emails} \\
  \midrule
  \multirow{5}{*}{Commits}
    & Identifiers        & \texttt{id}, \texttt{sha}, \texttt{repository} \\
    & Author fields      & \texttt{author\_login}, \texttt{author\_name}, \texttt{author\_email}, \texttt{author\_date} \\
    & Committer fields   & \texttt{committer\_login}, \texttt{committer\_name}, \texttt{committer\_email}, \texttt{committer\_date} \\
    & Git metadata       & \texttt{tree\_sha}, \texttt{parent\_shas}, \texttt{comment\_count}, \texttt{total\_changes} \\
    & Signing/verification & \texttt{signature}, \texttt{verified}, \texttt{verification\_reason}, \texttt{verified\_at} \\
  \bottomrule
  \end{tabularx}
\end{table}

\PP{Stratified Sampling Strategy.}
To efficiently capture a representative sample across this allocated range while keeping query volume manageable, we divided the approximately 230M identifier range into 23,000 equal intervals and queried \texttt{/users?since=\{id\}} at each interval's starting identifier, to give an approximate 1\%. This selected account blocks at regularly spaced points in the range. Each request returned up to 100 accounts after a selected ID. We repeated this at regular points across the ID range to include accounts created at different times while keeping the number of requests manageable. By including these different groups, we aimed to avoid overlooking accounts created during a particular period. This design had a maximum potential yield of 2.3M account records, which is about 1\% of the ID space, and we obtained profile metadata for 1,294,214 accounts. The remaining queries returned \texttt{NOT FOUND} and when we manually inspected a sample of these URLs via the web interface, GitHub returned a 404 page. This likely means that the accounts have been deleted or suspended~\cite{github_emu_docs}.

\PP{Sample Robustness and Expansion.}       
While our 1\% sample spans GitHub's full operational history (2008--2026)~\cite{britannica2024} and captures users across different signup periods, we note that this temporal coverage alone does not establish population representativeness. Treating the 1\% collection as an exploratory baseline, we expanded to 2\%, doubling coverage. We compared selected account and key proportions between the two collection stages.

To assess robustness, we expanded the sample to check whether the measured proportions changed as more accounts were included~\cite{provost1999efficient,riondato2015mining}. These proportions remained similar across the two samples. We stopped further expansion because more accounts would mean collecting additional repository and commit data which would require substantial resources, with limited expected benefit. We expanded the sample by retaining the initial 1\% sample and adding a further 1\% collected from starting points halfway between the original query points. The resulting 2\% sample therefore includes both the original accounts and the newly collected accounts.

Furthermore, we define active users as accounts with at least one publicly visible commit in a public repository, identified through the GitHub REST API. This definition captures public contribution at any point in an account’s history and does not require recent activity. Table~\ref{tab:sample_validation} shows that active-user prevalence and signing-key proportions remain stable across strata. While differences are statistically detectable due to our large sample sizes, the effect sizes are negligible, confirming that expanding the sample beyond 1\% does not materially shift population-level estimates.

\begin{table}
  \caption{Sampling robustness across strata. Effect sizes remain negligible across both dimensions despite large sample sizes.}
  \label{tab:sample_validation}
  \centering
  \small
  \setlength{\tabcolsep}{1pt}
  \begin{tabular}{l r r r}
    \toprule
    \multicolumn{4}{c}{\textit{Active-User Prevalence}} \\
    \midrule
    Stratum & $N$ & Active $N$ & \% \\
    \midrule
    1\% & 1,294,214 & 33,895  & 2.62 \\
    2\% & 2,737,649 & 81,540  & 2.98 \\
    \midrule
    \multicolumn{4}{l}{\textit{Statistics:} $\chi^2(1) = 408.38$, $p = 8.24\times10^{-91}$, Cram\'er's $V = 0.0101$} \\
    \midrule[\lightrulewidth]
    \multicolumn{4}{c}{\textit{Signing-Key Prevalence}} \\
    \midrule
    Stratum & Total Keys & Signing Keys & \% \\
    \midrule
    1\% & 98,590  & 10,229 & 10.38 \\
    2\% & 208,546 & 21,427 & 10.27 \\
    \midrule
    \multicolumn{4}{l}{\textit{Statistics:} $\chi^2(1) = 0.74$, $p = 0.39$, Cram\'er's $V \approx 0.002$} \\
    \bottomrule
  \end{tabular}
\end{table}

\subsection{Commit Metadata Collection} 
\label{sec:Commit}

We restricted commit metadata collection to active users (Table \ref{tab:sample_validation}), to focus on developers with tangible platform contributions. Collection proceeded via GitHub's REST API in two phases. 

%\PP{Repository discovery.} 
First, for each sampled account, we used \texttt{GET /users/\{username\}/repos} with \texttt{type=all} to retrieve the full public repository listing available through the endpoint~\cite{githublistuserrepos}. We requested 100 results per page and continued until a page returned fewer than 100 results. We retained all returned repositories, including forks, regardless of size, popularity, or activity. However, this procedure does not establish that every repository to which an account had contributed was recovered. This procedure may omit public repositories to which sampled users contributed, such as upstream repositories where they submitted pull requests from forks~\cite{githubcontributingtoaproject}. Also, private repositories were outside the scope of our collection as this is inaccessible via public APIs.

%\PP{Commit retrieval.} 
Second, for each public repository returned for a
sampled account, we queried \texttt{GET /repos/\{owner\}/\{repo\}/commits}
with the account's GitHub username as the \texttt{author} parameter. Filtering by author let us measure each sampled account’s signing behavior using the commits GitHub attributed to that account. We omitted \texttt{sha}, so each listing began at the repository's default branch, applying a consistent branch-selection rule across repositories~\cite{githubcommitsapi}. Commit requests used REST API version 2022-11-28. We requested up to 100 commits per page, starting with the first page, to reduce the number of requests. We treated a page with fewer than 100 commits as the end of the list. We also stopped if GitHub reported that the repository could not be found. In the latter case, the collector retained commits from any earlier pages. We retrieved detailed metadata through separate \texttt{GET} requests to commit URLs and recorded the fields listed in Table~\ref{tab:collected-data}. Our analyses therefore concern observed public commits retrieved through this procedure, rather than independently validated complete contribution histories.

\PP{Signature metadata and attribution.}
We recorded the author and committer separately because these fields can list different accounts. Being listed as a commit's author does not prove that the account signed it; someone else may have signed the commit. We used the signature information and verification result returned by GitHub. We did not re-verify older signatures ourselves using newer key information. GitHub can keep a commit marked as verified even if its signing key later expires or is revoked~\cite{github_commit_signature_verification}.

%\PP{Collection progress and stability.} Because commit retrieval is highly resource-intensive, we structured collection in stages. At the time of writing, the expanded collection had processed 71,694 of 81,540 eligible accounts (87.9\%), yielding 16,112,439 commit records across 874,198 repositories. This percentage measures account-level queue completion rather than partial commit recall; for every processed account, we collected the complete set of commits across all of its identified repositories. The pipeline processed the target account list through a randomized queue to spread retrieval across the selected list. Randomized ordering alone does not establish that the completed and pending accounts have the same characteristics, because activity volume may affect completion. We compared selected signing proportions between the initial and expanded collections as a descriptive stability check (Table~\ref{tab:robustness}). The expanded collection includes the initial collection, so these are overlapping cohorts. The measured proportions were similar across collection stages. We used the processed portion of the expanded collection for the commit-based analyses reported in this paper. 

\PP{Collection progress and stability.} 
Because collecting commit histories requires significant time and computing resources, we processed accounts in stages using a randomized queue. At the time of writing, we had completed 71,694 of 81,540 eligible accounts (87.9\%), gathering 16,112,439 commits across 874,198 repositories. For every completed account, we collected all commits from its identified repositories, meaning this 87.9\% reflects how many accounts are finished rather than partial commit data. However, accounts with heavy activity take longer to fetch, so a randomized queue alone does not guarantee that finished and unfinished accounts have identical traits. To check for bias, we compared signing proportions between the initial collection and the overlapping expanded collection (Table~\ref{tab:robustness}). Because these proportions remained stable across both stages, we use the 71,694 completed accounts for all commit-based analyses in this paper.

\begin{table}
\caption{Robustness check the initial 1\% sample and the expanded 2\% sample. Negligible absolute differences ($|\Delta| \le 1.21$\,pp) and trivial effect sizes ($|h| < 0.03$, where $0.2$ is the threshold for a small effect) confirm that the 1\% sample is representative and invariant to sample expansion.}
\label{tab:robustness}
\centering\small
\setlength{\tabcolsep}{3pt}
\begin{adjustbox}{max width=\columnwidth}
\begin{tabular}{@{}lrrrrrr@{}}
\toprule
Metric & 1\% & 2\% partial & $|\Delta|$ (pp) & $z$ & $p$ & $|h|$ \\
\midrule
\multicolumn{7}{@{}l}{\textit{All commits}} \\
Commits & 7,799,840 & 16,112,439 & --- & --- & --- & --- \\
Author accounts & 33,895 & 71,694 & --- & --- & --- & --- \\
Repositories & 414,151 & 874,198 & --- & --- & --- & --- \\
Signed commits & 23.74\% & 24.05\% & 0.32 & $-17.08$ & $2.16\times 10^{-65}$ & 0.0075 \\
Accounts ever signed & 89.19\% & 89.21\% & 0.02 & $-0.10$ & 0.919 & 0.0007 \\
Signed among signers & 24.24\% & 24.61\% & 0.37 & $-19.47$ & $1.89\times 10^{-84}$ & 0.0086 \\
\midrule
\multicolumn{7}{@{}l}{\textit{Excluding UI, author = committer}} \\
Commits & 6,146,794 & 12,617,006 & --- & --- & --- & --- \\
Author accounts & 28,789 & 61,154 & --- & --- & --- & --- \\
Repositories & 323,482 & 680,523 & --- & --- & --- & --- \\
Signed commits & 5.32\% & 5.22\% & 0.10 & $9.36$ & $7.69\times 10^{-21}$ & 0.0046 \\
Accounts ever signed & 5.25\% & 5.15\% & 0.10 & $0.63$ & 0.530 & 0.0045 \\
Signed among signers & 30.42\% & 31.63\% & 1.21 & $-22.01$ & $2.41\times 10^{-107}$ & 0.0262 \\
\midrule
UI commit fraction & 20.21\% & 20.63\% & 0.42 & $-23.82$ & $1.91\times 10^{-125}$ & 0.0104 \\
\bottomrule
\end{tabular}
\end{adjustbox}
\par\smallskip
\begin{minipage}{\columnwidth}
\footnotesize\raggedright
Two-sided pooled $z$-tests ($z = \text{rate}_1 - \text{rate}_2$); $p$-values are nominal due to overlapping samples. For subsets excluding web UI commits, we require matching author/committer logins and exclude \texttt{web-flow}.% (0 \texttt{[bot]} accounts found), and omit missing author logins (1,590 in 1\%; 4,209 in 2\%) from rates/tests while keeping them in raw counts.
\end{minipage}
\end{table}

%% file: sections/5.data-analysis.tex
\section{Data Analysis}
\label{sec:data-analysis}

\PP{Commit Filtering.}
Table~\ref{tab:collected-data} summarizes all collected fields. We used account fields to filter for active users and characterize our sample. To determine whether a commit was signed, we checked for a non-null \texttt{signature} field. We then examined GitHub's \texttt{verified} and \texttt{verification\_reason} fields to distinguish between commits that merely had a signature attached and those that GitHub successfully verified.

Next, we partitioned commits into three mutually exclusive authoring mechanisms (Table~\ref{tab:signing_breakdown}): \textit{Web}, \textit{E2E}, and \textit{Other}. We identified \textit{Web} commits made via the GitHub web interface by checking for \texttt{committer\_login=web-flow} and \texttt{committer\_name=GitHub}. However, simply excluding this GitHub-controlled path does not guarantee that the remaining commits were E2E authored by a single developer. To establish a stronger heuristic for our E2E commit analysis (\S\ref{subsection:signing_adoption}--\S\ref{subsec:signing-key-results}), we defined \textit{E2E} commits as non-\texttt{web-flow} commits where the author and committer logins matched. All remaining commits, including those with mismatched logins or a missing login, were classified as \textit{Other}. The E2E rule aligns with the standard workflow in which developers sign their own commits locally using \texttt{-S} before pushing~\cite{github_commit_signature_verification}. After applying this filter, we checked all E2E commits and found zero accounts ending in \texttt{[bot]} (the suffix GitHub uses for bot accounts~\cite{wyrich2021bots}) in either the author or committer fields. This confirms that our matching-login rule automatically filtered out bot-pushed commits, as well as commits where a maintainer merged another contributor's code under a different username.

While matching logins for author and committer and the absence of bots increase confidence that a commit came from a specific individual, this heuristic has limitations. An automated script or CI/CD pipeline configured with a developer's credentials can still put the same login in both fields~\cite{goyal2018identifying}, and another person could sign a commit if credentials or keys are shared~\cite{torres2016omitting}. Because of this, we report signing activity as observed per GitHub account rather than claiming with certainty that the account owner personally held the signing key.

\PP{Key Attribution.}
Because these commit metadata as detailed in Table~\ref{tab:collected-data} cannot tell us whose key produced a signature, we next tie signatures to the public keys registered on GitHub. To do so, we first connected each uploaded key to a GitHub account by matching its \texttt{user\_id} to the \texttt{id} in the collected account profiles. This provided the account login associated with each uploaded key. We then recovered key fingerprints from commit signatures and compared them with fingerprints obtained from the uploaded public keys. Matches required the same signature format and full fingerprint. Matching was performed against the full collected key inventory, rather than being restricted to keys uploaded by the commit's author, so that we could identify signatures made with keys registered to other accounts. Matches based only on a shorter GPG key ID were kept separate as possible matches, because OpenPGP key IDs are not guaranteed to be unique~\cite{rfc9580}. 

\PP{Observation Window.} To bound our observation windows and temporal calculations throughout the paper, we used April~12, 2026 at 23:00:49 UTC—the timestamp of the last crawled commit at the time of data collection—as the end of our coverage period.

%% file: sections/7.results.tex
\section{Results}
\label{sec:results}
To answer our research questions, we characterize the overall dataset using all commits (\S\ref{subsection:overview}) and analyze the restricted E2E subset in \S\ref{subsection:signing_adoption}--\S\ref{subsec:signing-key-results}.

\subsection{Dataset Overview}
\label{subsection:overview}
At first glance, commit signing appears pervasive: 89.21\% of active users (63,961 of 71,694) have authored at least one signed commit, and 24.05\% of all commits (3,874,732 of 16,112,439) carry a signature (Table~\ref{tab:robustness}). However, Table~\ref{tab:signing_breakdown} shows that this signing is driven almost entirely by GitHub's web interface, which signs automatically (96.41\% signed) and accounts for 82.69\% (3,203,969 of 3,874,732) of all signed commits in our dataset. In contrast, E2E commits are rarely signed (5.22\%; 658,217 of 12,617,006). Although web-interface signing is not our primary focus, auditing it for completeness revealed 119,433 of 3,323,402 unsigned web commits (3.59\%). While most predate GitHub's introduction of automatic web signing in late 2017~\cite{afzali2020towards}, 28,563 of 119,433 (23.92\%) occurred after 2017 (Table~\ref{tab:webflow_audit} in Appendix~\ref{others}).

\begin{table}
\centering
\caption{Commit signing by authoring mechanism.}
\label{tab:signing_breakdown}
\begin{threeparttable}
\small
\setlength{\tabcolsep}{4pt}
\begin{tabular}{lrrrr}
\toprule
\textbf{Mechanism}\tnote{a} & \textbf{Commits} & \textbf{\% all} & \textbf{Signed} & \textbf{\% signed} \\
\midrule
E2E                         & 12,617,006 & 78.31  & 658,217   & 5.22  \\
Web (\texttt{web-flow})     & 3,323,402  & 20.63  & 3,203,969 & 96.41 \\
Other                       & 172,031    & 1.07   & 12,546    & 7.29  \\
\midrule
\textbf{Total}              & 16,112,439 & 100.00 & 3,874,732 & 24.05 \\
\bottomrule
\end{tabular}
\end{threeparttable}
\end{table}

\subsection{Adoption}
\label{subsection:signing_adoption}

Looking strictly at the E2E commits, we found that 61,154 of the 71,694 authors had at least one E2E commit. Of these authors, only 5.15\% (3,151 / 61,154) ever signed an observed E2E commit. Moreover, developers rarely start signing when they create an account. Of the 3,151 accounts with at least one observed signed E2E commit, 3,132 had a valid account creation date and a first observed signature dated on or after account creation. We excluded one of these accounts because both signed and unsigned commits shared their earliest observed timestamp. Of the remaining 3,131 accounts, only 453 (14.47\%) signed their very first E2E commit. The other 2,678 accounts are late adopters, whose first observed E2E commits were unsigned.

\begin{figure}
  \centering
  \includegraphics[width=0.80\linewidth]{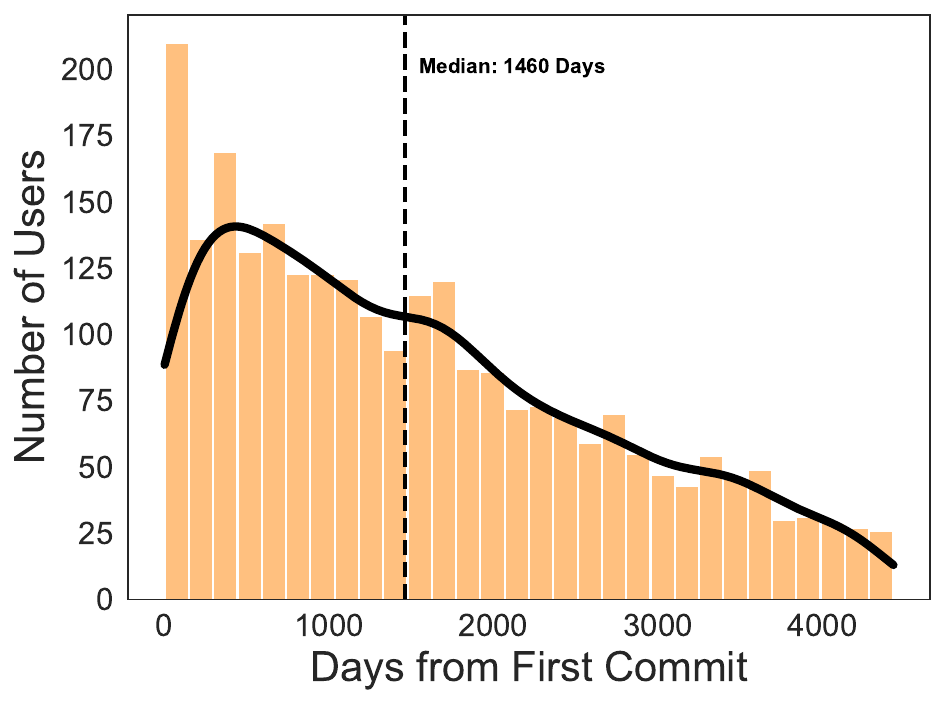}
  \caption{Days from first E2E commit to first signed E2E commit among late adopters.}
  \label{fig:adoption_latency_days}
  \Description{A cumulative distribution plot showing the number of days from first observed E2E commit to first signed commit for late adopters.}
\end{figure}

\begin{figure}
  \centering
  \includegraphics[width=0.80\linewidth]{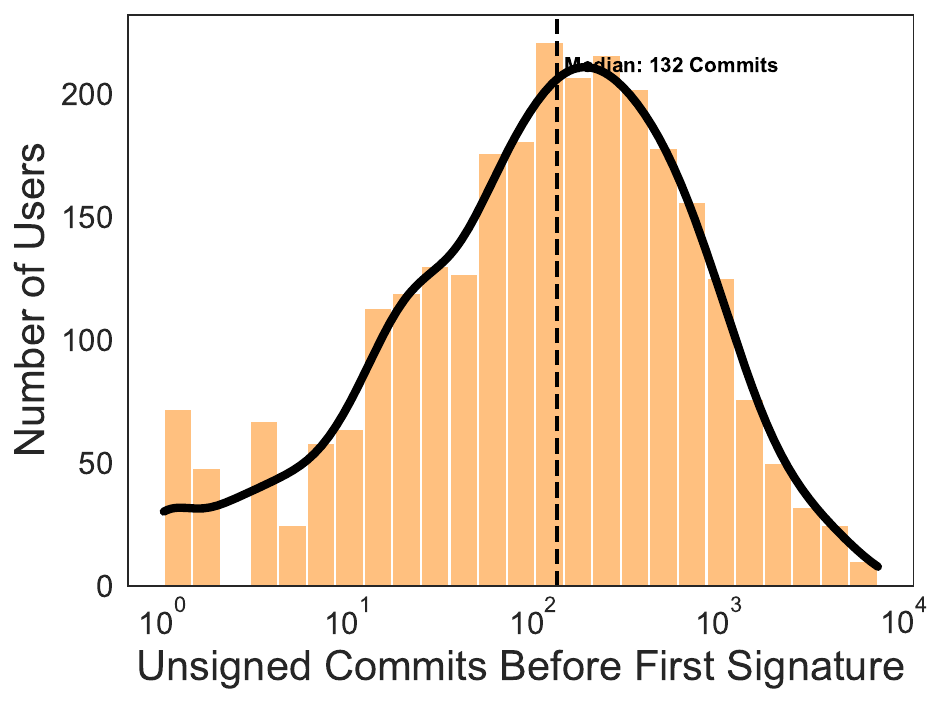}
  \caption{Unsigned E2E commits authored before first signed E2E commit among late adopters.}
  \label{fig:adoption_latency_commits}
  \Description{A distribution plot showing the frequency of unsigned E2E commits made prior to the first signed commit.}
\end{figure}

Across all 3,131 signers, the median time to first signature is 1,112 days after the first observed commit and 2,154 days after account creation. This figure includes the early signers, so it understates the delay for late adopters. Among late adopters alone, the median delay between their first observed commit and their first signed commit is 1,460 days (IQR: 623--2,584), or roughly four years. As a result, late adopters accumulate a large amount of unsigned work before signing for the first time, with a median of 132 unsigned commits (IQR: 32--402) (Figure~\ref{fig:adoption_latency_days} and Figure~\ref{fig:adoption_latency_commits}). Table~\ref{tab:signing-by-age} further corroborates this trend across account age brackets: the share of accounts that ever signed an observed E2E commit rose across age groups, with a statistically significant correlation (Spearman $\rho = 0.1993$, $p < 0.001$).

\begin{table}
  \centering
  \caption{E2E commit signing adoption by account age.}
  \label{tab:signing-by-age}
  \scriptsize
  \setlength{\tabcolsep}{4pt}
  \begin{tabular}{@{}lrrr@{}}
  \toprule
  Account age (years) & Users & Ever signed E2E & Signing rate \\
  \midrule
  $\le 1$ & 2,226 & 16 & 0.72\% \\
  1--2 & 10,065 & 93 & 0.92\% \\
  2--5 & 20,266 & 394 & 1.94\% \\
  5--10 & 15,930 & 987 & 6.20\% \\
  $>$10 & 12,644 & 1,656 & 13.10\% \\
  \bottomrule
  \multicolumn{4}{@{}p{0.82\columnwidth}@{}}{{\scriptsize Note: Excludes 23 accounts (5 E2E signers) with missing or invalid account creation dates ($N = 61{,}131$ users; $3{,}146$ signers).}}
\end{tabular}
\end{table}

\begin{mytakeaway}
E2E commit signing is rare and most adopters begin years after creating their account, leaving a large backlog of unsigned commits first, which explains why E2E signing rates rise steadily with account age.
\end{mytakeaway}

\subsection{Post Adoption}
\label{subsection:signing_after_adoption}

\begin{figure}
  \centering
  \includegraphics[width=0.85\linewidth]{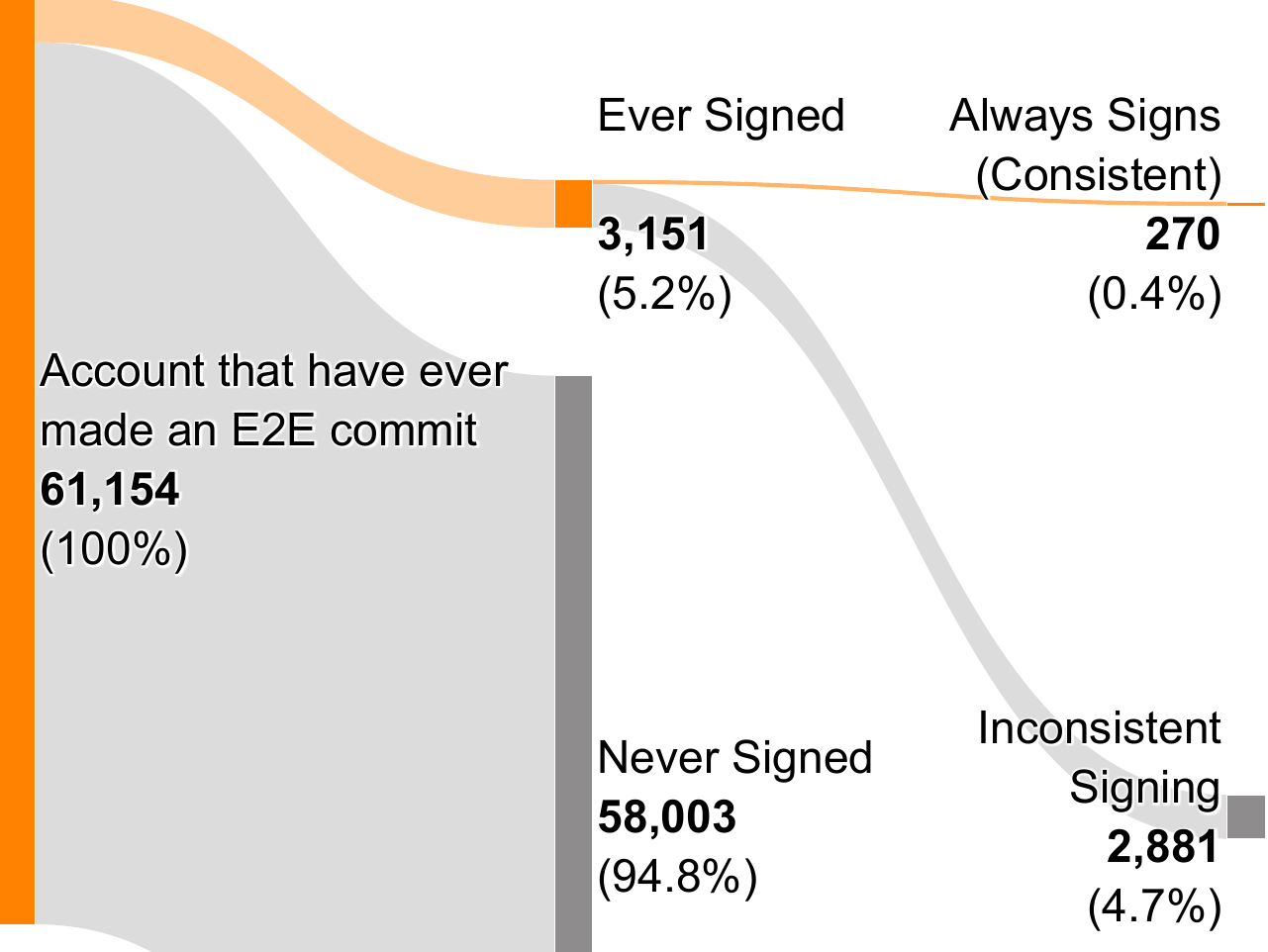}
  \caption{Commit signing habits among E2E accounts.}
  \label{fig:sankey_nonui}
  \Description{Sankey diagram showing author commit signing breakdown: 61,154 total accounts (100\%), 58,003 never signed (94.8\%), 3,151 ever signed (5.2\%), 2,881 signed inconsistently (4.7\%), and 270 consistently signed all commits (0.4\%).}
\end{figure}

Next, we look at how often developers sign their commits and whether they eventually stop. Figure~\ref{fig:sankey_nonui} breaks down these signing habits for all E2E authors.

\begin{figure}
  \centering
  \includegraphics[width=0.80\linewidth]{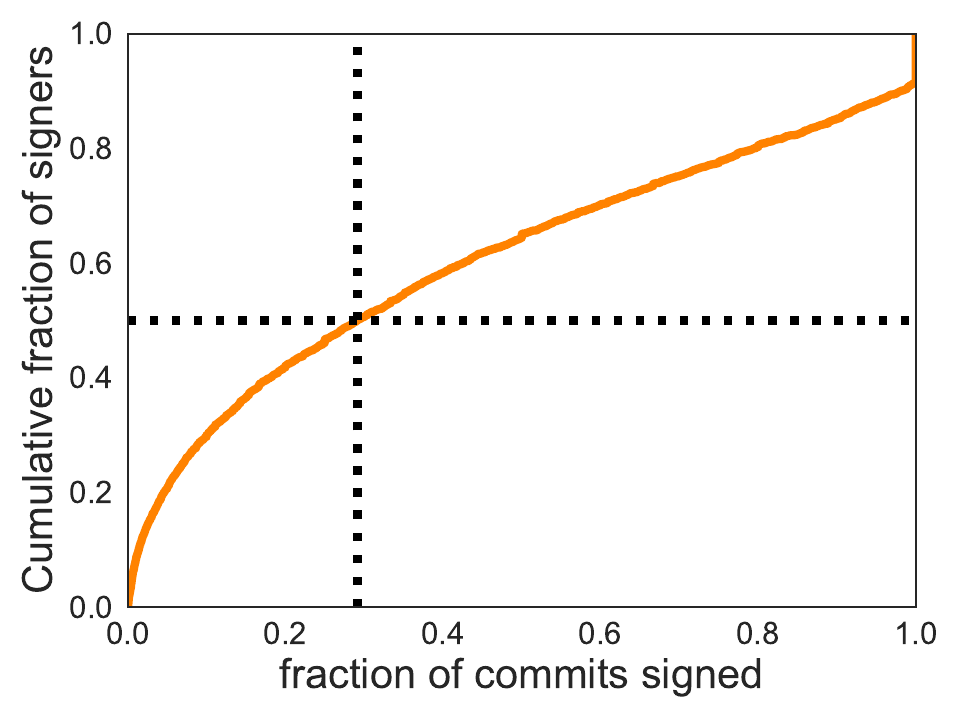}
  \caption{CDF of per-user signing rates, computed over E2E commits only to show signing consistency.}
  \label{fig:cdf}
  \Description{Cumulative distribution function plot showing the distribution of per-user commit signing rates.}
\end{figure}

\PP{How Consistent Is E2E Signing?} Among authors who produced at least one E2E signature, we found a total of 2,081,235 E2E commits, of which 658{,}217 are signed---an overall signing rate of 31.63\%. Thus, even among authors with at least one E2E signature, roughly one in three observed E2E commits carries a cryptographic signature. Shifting from aggregate commits to individual accounts, Figure~\ref{fig:cdf} shows that the median E2E signer signs only 29.13\% of their commits. Only 22.56\% (711 users) sign at least 75\% of their commits, and just 8.57\% (270 users) sign every observed commit, while the remaining 91.43\% sign only a subset of their commits.

\PP{Do Developers Subsequently Lapse?}
Beyond these overall signing rates, we examined whether developers continued contributing unsigned commits after their last observed signature. In the full-history analysis, 1,072 of 3,151 E2E signers (34.02\%) produced unsigned commits after their last observed signed commit, with a median of 28 subsequent unsigned commits (75th percentile: 120). This pattern also occurs among developers with high overall signing rates. Among the 711 users who signed at least 75\% of their observed E2E commits, 64 (9.00\%) produced unsigned commits after their last observed signed commit, with a median of 4 subsequent unsigned commits (75th percentile: 7).

\PP{Does the Pattern Remain Over Fixed Observation Periods?} Because accounts created earlier have longer observation windows, unadjusted lapse rates might simply reflect the passage of time rather than a true behavioral pattern. To control for this, we repeated the analysis over fixed periods of 90, 180, and 365 days after each user's first observed signed E2E commit. This check started with 3,151 E2E signers. For each period, we included users whose full window ended by the cutoff. Percentages were calculated among users who made at least one further observed commit during that window. For each user and window, we identified the last signed commit and counted unsigned commits that followed it before the window ended. We also applied a stricter rule requiring at least five such commits, with the latest at least 30 days after the last signed commit. Table~\ref{tab:fixed-followup} shows that unsigned contributions appeared in all three windows under both rules. Among users who made another commit, 9.67\% to 19.24\% met the stricter rule.

\begin{table}
\centering
\caption{Unsigned commit activity and lapsing rates across fixed follow-up windows.}
\label{tab:fixed-followup}
\small
\setlength{\tabcolsep}{2pt}
\begin{tabular}{@{}rrrrr@{}}
\toprule
Days & Eligible & \shortstack[r]{With\\follow-up} & \shortstack[r]{At least 1\\unsigned} & \shortstack[r]{At least 5\\(30-day rule)} \\
\midrule
90 & 3,086 & 2,771 & 609 (21.98\%) & 268 (9.67\%) \\
180 & 2,956 & 2,745 & 672 (24.48\%) & 362 (13.19\%) \\
365 & 2,595 & 2,464 & 743 (30.15\%) & 474 (19.24\%) \\
\bottomrule
\end{tabular}
\end{table}

\PP{Do Accounts Return to Signing After Making Unsigned Commits?} We also checked whether users signed again after making unsigned commits. During the 365-day period, 1,448 users made at least one unsigned commit after their first observed signed commit. Of these, 1,063 (73.41\%) later signed again within the same period. Returning to signing and later contributing unsigned commits can occur in the same history. A user could make an unsigned commit, return to signing, and then make further unsigned commits after their last signature in the period. Among the 743 users with unsigned commits after their last signed commit during the 365-day period, 416 (55.99\%) made such commits in a repository where they had already signed during that period.

%These results show that signing can be intermittent, including within the same repository. They do not establish that users stopped signing permanently.

\PP{Do accounts who sign their first commit maintain it long-term?} Of the 453 users who signed their first observed E2E commit, 83 (18.32\%) finished with an unsigned latest E2E commit by the observation cutoff. Among these 83 users, the median time elapsed between their final observed E2E commit and the cutoff was 170 days (IQR 113--336). Furthermore, for 79.52\% of these users, their last observed commit occurred within 365 days of the cutoff.

\PP{Does Signing Lapse Vary with Account Age?} Table~\ref{tab:lapse-by-age} shows higher unadjusted rates of unsigned contributions after the last observed signature among older accounts. However, these rates do not account for differences in follow-up time or activity and do not establish abandonment.

\begin{table}
\centering
\caption{Unadjusted proportions of E2E signers with unsigned commits after their last observed signed commit, by account age at the cutoff (Spearman $\rho = 0.0798$, $p = 7.46 \times 10^{-6}$).}
\label{tab:lapse-by-age}
\small
\setlength{\tabcolsep}{4pt}
\begin{tabular}{@{}lrr@{}}
\toprule
Account age (years) & E2E signers & Lapse rate \\
\midrule
$\le$1 & 16 & 6.25\% \\
1--2 & 93 & 18.28\% \\
2--5 & 394 & 28.17\% \\
5--10 & 987 & 33.13\% \\
$>$10 & 1,656 & 37.08\% \\
\bottomrule
\multicolumn{3}{@{}p{0.82\columnwidth}@{}}{{\scriptsize \textit{Note:} Excludes 5 E2E signers with missing or invalid account creation dates ($n = 3{,}146$).}} \\
\end{tabular}
\end{table}

We also examined account age at the first observed signature within the 90-, 180-, and 365-day periods, excluding 13--14 active signers per window with missing or invalid account creation dates. We accounted for the log-transformed counts of later commits and repositories and for the calendar year of the first observed signature. Account age was not statistically associated with making at least one unsigned commit after the period's last signed commit in any of the three periods ($n = 2{,}758, p = 0.7153$ at 90 days; $n = 2{,}731, p = 0.2748$ at 180 days; and $n = 2{,}450, p = 0.3731$ at 365 days). In the 365-day model ($n = 2{,}450$), the odds ratio per additional account year was 1.011 (95\% CI: 0.9866--1.037; $p = 0.3731$). When we required at least five unsigned commits, with the latest at least 30 days after the last signed commit, the age odds ratios were 1.024 at 90 days ($p = 0.2115$), 1.024 at 180 days ($p = 0.1395$), and 1.016 at 365 days ($p = 0.2904$). None of these six adjusted models showed a statistically significant relationship between account age and signing lapse.

\PP{Is E2E Signing Consistent Across Different Repositories?} Having shown that E2E commit signing is rare, we examined how consistently developers sign across their observed E2E contributions. As detailed in Figure~\ref{fig:sankey_nonui}, 91.43\% (2,881) of the 3,151 users who ever signed an E2E commit had inconsistent signing records.

At the repository level, only 14.54\% (458) of E2E signers had at least one signed commit in every repository they contributed to, while the vast majority signed in some repositories but not others (2{,}693; 85.46\%). The difference in repository breadth between these groups is statistically significant (Mann--Whitney $U = 121{,}495$, $p < 0.001$). These findings indicate that signing is not a uniform behavior across developers' contributions. To characterize this variation in more detail, we examined each user's signing behavior within individual projects. The vast majority of E2E signers (2,990 of 3,151) contributed to two or more repositories. Among these users, 4.98\% (149) signed the same fraction of commits in every repository, whereas 95.02\% (2,841) had different signing fractions across repositories (median within-account standard deviation = 0.3652).

We then examined whether this variation was associated with users' contribution activity within repositories. We grouped 79,269 user--repository pairs from E2E signers by commit activity level, comparing pairs with 1--10 commits to those with 11 or more. Although the difference is statistically significant (Mann--Whitney $U = 674{,}453{,}488$, $p = 2.43 \times 10^{-56}$), the practical effect is negligible (Cliff's $\delta = -0.057$). The mean signing rate was 29.11\% for the 1--10 commit group and 30.35\% for the 11-or-more group. The median signing rate was 0\% in both groups.

\begin{figure}
  \centering
  \includegraphics[width=0.80\linewidth]{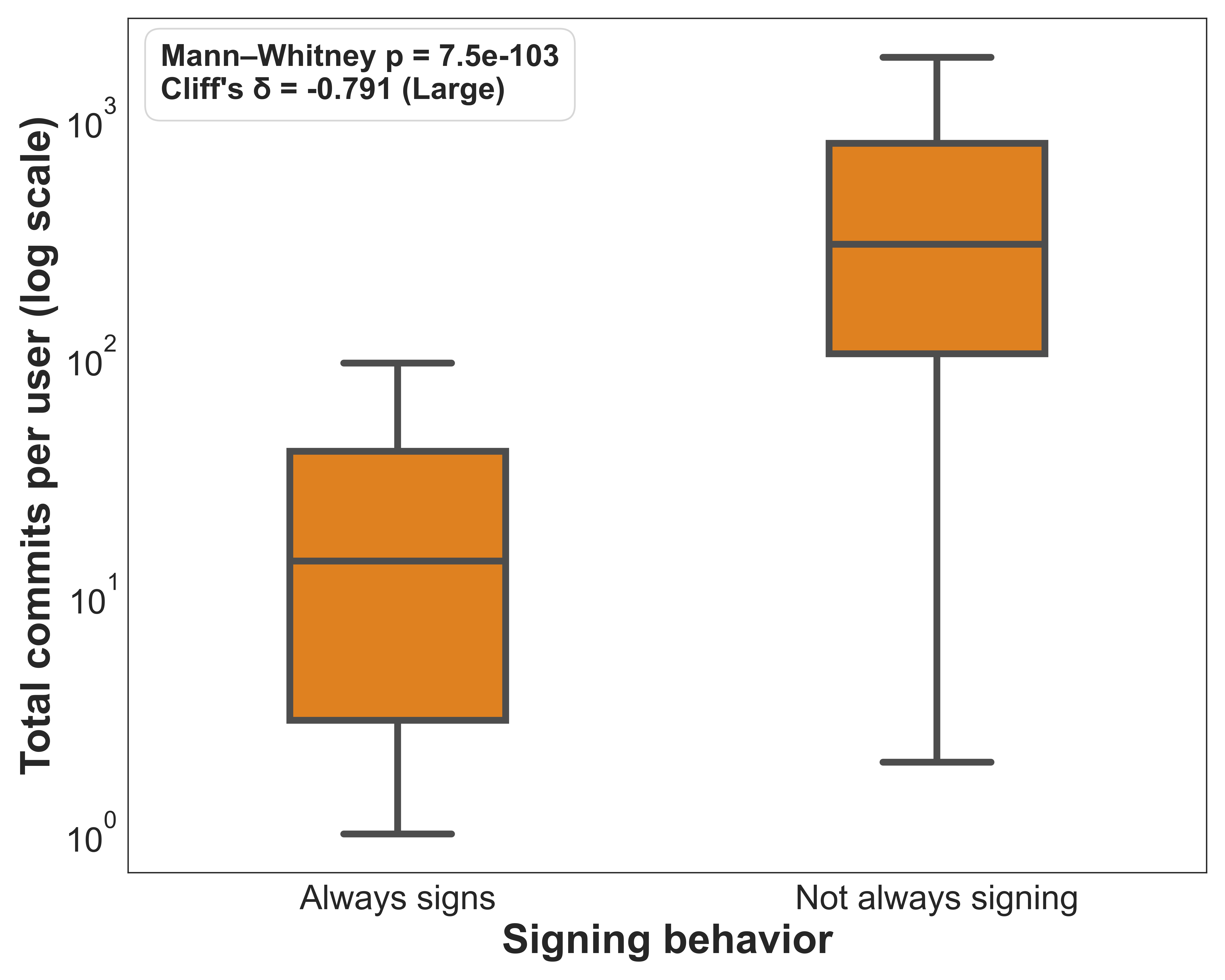}
  \caption{Distribution of observed E2E commit counts per user on a logarithmic scale, comparing always-signers with intermittent signers.}
  \label{fig:consistency}
  \Description{Logarithmic distribution plot comparing commit counts between always-signers and intermittent signers, showing that always-signers cluster heavily at lower commit volumes.}
\end{figure}

\PP{How Does Signing Consistency Relate to Developer Activity?} We found that accounts that sign every observed E2E commit have substantially smaller observed commit histories than intermittent signers (median: 14 vs. 300 commits). Also, 45.56\% of these always-signers have at most 10 commits (Figure~\ref{fig:consistency}). Given the highly skewed commit counts, we used a Mann--Whitney $U$ test~\cite{mcknight2010mann}, finding a significant difference ($p = 7.49 \times 10^{-103}$) with a large effect size (Cliff's $\delta = -0.791$)~\cite{meissel2024using}.

\begin{mytakeaway}
Even among E2E signers, signing is inconsistent, largely intermittent, with accounts frequently lapsing into unsigned work and later returning to signing regardless of account age. Perfect consistency is rare and heavily concentrated among accounts with smaller commit histories.
\end{mytakeaway}

\subsection{Signing Keys}
\label{subsec:signing-key-results}

Across the 3,151 author accounts with at least one E2E signed commit, we found 658,217 total signed commits. This subsection analyzes these commits and their associated signing keys in detail.

\PP{How many signing keys does an author account have?}
We grouped signed commits by author account and counted the distinct key identifiers found in their signatures. Repeated use of the same identifier within an account was counted only once. We found that most accounts had one identifier (1,896 of 3,151; 60.17\%), while 1,255 accounts (39.83\%) had multiple identifiers (Table~\ref{tab:signing-key-counts} in Appendix~\ref{others}). Every account had at least one usable identifier. These counts describe observed identifiers: some GPG signatures provided only a short key ID rather than a full fingerprint.

\PP{Can signatures be connected to uploaded public keys?} Because we do not have access to developers' private keys, we matched signature key identifiers against our collected public-key inventory and cross-tabulated the results with GitHub's push-time verification records (Table~\ref{tab:registry_matching}). Most GPG and SSH signatures matched an uploaded key by exact fingerprint, whereas CMS/S/MIME signatures had no matches because GitHub does not expose X.509 certificates in user key registries. Notably, every exact-matched signature marked \texttt{valid} by GitHub matched a public key registered to the commit author's own GitHub account. A small subset of GPG signatures matched an uploaded key only by short key ID; we treat these as possible matches and exclude them from key-size analyses because OpenPGP key IDs are not guaranteed to be unique~\cite{rfc9580}. Cross-tabulating against \texttt{verification\_reason} shows that while most unmatched signatures failed at push time with \texttt{unknown\_key} (keys never uploaded to GitHub), 4,137 \texttt{unknown\_key} commits now match a key registered to the author's account (keys uploaded after those commits were pushed). Conversely, nearly a third of currently unmatched signatures (30,459 of 91,570) were verified as \texttt{valid} when pushed---demonstrating that tens of thousands of previously verified commits were signed with keys later deleted from GitHub, an operational blind spot we examine in \S\ref{sec:discussion}.

\begin{table}
\centering
\caption{Public-key registry matching and push-time verification outcomes for signed E2E commits.}
\label{tab:registry_matching}
\resizebox{\columnwidth}{!}{%
\begin{tabular}{lrrrr}
\toprule
\multicolumn{5}{l}{\textbf{Part A: Snapshot Registry Matching by Signature Format}} \\
\midrule
\textbf{Format} & \textbf{Signed (\%)} & \textbf{Exact} & \textbf{Short ID} & \textbf{Unmatched} \\
\midrule
GPG & 569,985 (86.60) & 469,921 & 25,921 & 74,143 \\
SSH & 86,287 (13.11) & 70,805 & 0 & 15,482 \\
CMS/S/MIME & 1,945 (0.30) & 0 & 0 & 1,945 \\
\midrule
Total & 658,217 (100.00) & 540,726 & 25,921 & 91,570 \\
Share (\%) & 100.00 & 82.15 & 3.94 & 13.91 \\
\midrule
\multicolumn{5}{l}{\textbf{Part B: Push-Time Verification (\texttt{verification\_reason}) vs. Registry}} \\
\midrule
\textbf{Reason} & \textbf{Total (\%)} & \textbf{Exact} & \textbf{Short ID} & \textbf{Unmatched} \\
\midrule
\texttt{valid} & 576,010 (87.51) & 521,089 & 24,462 & 30,459 \\
\texttt{unknown\_key} & 61,547 (9.35) & 4,151 & 9 & 57,387 \\
\texttt{unverified\_email} & 12,558 (1.91) & 8,962 & 1,263 & 2,333 \\
\texttt{bad\_email} & 5,958 (0.91) & 5,908 & 38 & 12 \\
\texttt{bad\_cert} & 1,379 (0.21) & 0 & 0 & 1,379 \\
\texttt{not\_signing\_key} & 622 (0.09) & 473 & 149 & 0 \\
\texttt{invalid} & 121 (0.02) & 121 & 0 & 0 \\
\texttt{gpgverify\_error} & 22 ($<$0.01) & 22 & 0 & 0 \\
\bottomrule
\end{tabular}}

\parbox{\columnwidth}{\scriptsize
Note: Of the 540,726 exact matches, 540,712 (including all 521,089 exact-matched \texttt{valid} commits) matched a key uploaded by the commit's author; the remaining 14 matched another sampled account's key and were rejected as \texttt{unknown\_key}.}
\end{table}

\PP{Do accounts move to new signing keys or return to earlier keys?}
For the 1,255 accounts with multiple key identifiers, we ordered signed commits by committer date, using the author date when the committer date was missing. We found that 679 accounts (54.10\%) used keys without returning to an earlier one, following patterns such as A--B--C. Another 414 accounts (32.99\%) returned to an earlier key, following patterns such as A--B--A. We required full fingerprints and a clear time order to classify accounts. This left 162 accounts (12.91\%) unclassified. All 162 lacked a full fingerprint for at least one signature.

\PP{Do accounts use different signing keys for different repositories?}
We tried to see whether changing repositories could explain why accounts returned to earlier keys. Of the 1,255 groups with multiple key identifiers, we excluded 46 with signed commits in only one repository and 162 with incomplete fingerprint information. Of the 1,047 eligible accounts, 289 (27.60\%) used one key per repository and 758 (72.40\%) used multiple keys in at least one repository. Across both groups, 975 accounts (93.12\%) reused a key across repositories.

Among the 1,047 eligible accounts with a clear return pattern, multiple keys in a repository were more common among accounts that returned to an earlier key (357 of 402, 88.81\%) than among those with no observed return (401 of 645, 62.17\%). The gap was 26.64 percentage points (Pearson $\chi^2(1, N = 1{,}047) = 87.92$, $p = 6.818 \times 10^{-21}$, Cramér's $V = 0.290$). The patterns partly overlap because returning to a key within a repository requires using multiple keys there.

We separately examined all 414 accounts that returned to an earlier key in the preceding analysis, including accounts with signed commits in only one repository. Of these, 261 (63.04\%) returned within at least one repository. Another 45 (10.87\%) used one key per repository and returned across repositories. The remaining 108 (26.09\%) used multiple keys in at least one repository, with returns visible only in their combined history. For example, A--B in one repository followed by A in another forms an overall A--B--A pattern without a return within either repository.

\PP{When accounts resume signing after making unsigned commits, do they use the same key?}
For each of the 1,063 accounts that resumed signing within 365 days of its first observed signed E2E commit, we compared the signing keys immediately before and after its first run of unsigned commits. We ordered commits by author timestamp and required a single full fingerprint at each signed boundary. We observed the same fingerprint for 781 accounts (73.47\%) and a different fingerprint for 124 (11.67\%). Of the 124 accounts that changed keys, 113 used a fingerprint first observed at resumption and 11 returned to a previously observed fingerprint. The remaining 158 accounts (14.86\%) were unresolved because of insufficient fingerprint evidence (125), tied author timestamps (28), or both (5).

\PP{Are uploaded keys weak because of their size, or do they use deprecated algorithms?}
We classified key components with an estimated RSA-equivalent size below 2,048 bits as weak, following NIST's benchmark where 2,048-bit RSA corresponds to approximately 112 bits of security strength~\cite{barker2018recommendation}. Our public-key dataset contained 208,546 entries collected from GitHub. Because some GPG entries contained primary keys and subkeys assessed separately, counting each component once per account and category produced 225,865 component records (including authentication keys and GPG components serving purposes other than signing). Table~\ref{tab:uploaded-key-sizes} shows the breakdown.

We could not assess the size-based strength of 135 GPG components (0.06\% of all records). These unknown assessments—consisting entirely of Elgamal encryption components—remain included in the GPG and overall totals. OpenPGP RFC~9580 deprecates Elgamal keys and prohibits generating new ones or encrypting with them; because these are encryption components, their presence does not indicate Elgamal commit signatures.

Similarly, we found 143 DSA components (0.06\% of all component records).  NIST FIPS~186-5 no longer approves DSA for generating new digital signatures. Of these 143 components, 61 (42.7\%) were also weak by our key-size threshold, while the remaining 82 (57.3\%) met that threshold, demonstrating that a component can pass a size test while still relying on a deprecated algorithm.

\begin{table}
\centering
\caption{Uploaded keys assessed for weak key sizes.}
\label{tab:uploaded-key-sizes}
\scriptsize
\setlength{\tabcolsep}{4pt}
\begin{tabular}{@{}lrrr@{}}
\toprule
Key category & Components & Weak & Weak (\%) \\
\midrule
GPG                & 33,843  & 207 & 0.61 \\
SSH authentication & 187,119 & 226 & 0.12 \\
SSH signing        & 4,903   & 0   & 0.00 \\
\midrule
Total              & 225,865 & 433 & 0.19 \\
\bottomrule
\end{tabular}
\end{table}

\PP{Do weak keys appear in commit signatures, and how recently?}
We matched 16 of the 433 weak uploaded key components (3.70\%) to commit signatures. All 16 were GPG components. These matches covered 1,553 of the 658,217 signed commits (0.24\%), associated with 16 of the 3,151 author accounts (0.51\%). Weak keys continued to appear in commits dated 2025. We found 218 signed commits from that year that matched weak keys, with the latest dated December 4, 2025. For 12 of the 16 affected accounts, their latest observed signed commit also matched a weak key. For eight of these 12 accounts (66.67\%), that commit was dated 2025. For the remaining four, it was dated 2024 or earlier. Table~\ref{tab:latest-weak-key-years} in Appendix~\ref{others} shows this distribution.

\PP{Do deprecated algorithms appear in commit signatures, and how recently?}
We matched 20 of the 143 uploaded DSA key components (13.99\%) to commit signatures. These matches covered 4,516 of the 658,217 signed commits (0.69\%), associated with 19 of the 3,151 author accounts (0.60\%). We also checked algorithm information directly in GPG, SSH, and CMS/S/MIME signatures. DSA and SHA-1 appeared only in GPG signatures. Among the 569,985 GPG signed commits, 5,088 (0.89\%) used DSA, and 14,866 (2.61\%) used SHA-1 as the signature hash. These counts include commits without an uploaded-key match because we obtained the algorithm information from the signatures themselves.

Table~\ref{tab:signing-issues-by-year} in Appendix~\ref{others} summarizes the yearly counts of signatures matching weak key components, DSA signatures, and SHA-1 signatures. We continued to see DSA signatures into 2026, with the latest on March 11, 2026. SHA-1 signature counts generally declined after their peak, though we still observed SHA-1 as recently as December 4, 2025. Because a single signature can combine multiple issues, these counts overlap and should not be summed.

\PP{Does the same fingerprint appear across different accounts?} We recovered full fingerprints for 626,056 of the 658,217 signature records. We then kept records with a known GitHub author and counted each fingerprint once per signature format, yielding 6,115 distinct fingerprints. This count included fingerprints with and without a match to an uploaded public key. Of these, 3 (0.05\%) appeared on commits attributed to more than one GitHub account.

\begin{mytakeaway}
60\% of E2E signers use a single key and keep it when resuming after lapses, while others change or return to older keys across repositories, and a few key fingerprints even span multiple accounts. Most signatures match uploaded public keys, though a small number still use weak key sizes or outdated algorithms in recent commits.
\end{mytakeaway}

%% file: sections/6.limitation.tex
\section{Limitations}
\label{sec:limitations}

Several limits bound our measurement study. First, we acknowledge that GitHub’s API rate limits prevented exhaustive enumeration of all platform users, and that bucket-based sampling can miss certain account distributions. Furthermore, because GitHub IDs are assigned monotonically based on account creation date rather than in a strictly gapless sequence, ID gaps and unavailable profiles inevitably influence which accounts enter the sample. To mitigate these constraints, we employed progressive stratified sampling~\cite{provost1999efficient}, using the \texttt{since} parameter to retrieve blocks of up to 100 consecutive accounts following regularly spaced starting IDs. Capturing consecutive blocks preserves local account distributions and reduces the risk of overlooking specific cohorts. Moreover, because our expanded collection incorporates the initial sample, the stability observed across strata demonstrates consistency as the sample grew. Although this offers limited evidence regarding absolute representativeness, it strongly indicates that our findings are robust and not mere artifacts of sample size or identifier gaps.

We also recognize that restricting our analysis to commits reachable from the default branch of returned public repositories leaves private and enterprise activity unobserved, and can omit external repositories to which an account contributed as well as commits reachable only from non-default branches. We also note that these omitted commits can shift an account's first observed signature, make an account with signed commits appear to have none, or artificially create or hide gaps in signing activity. Nevertheless, focusing on default branches of publicly accessible repositories establishes a consistent, reproducible baseline for measuring standard open-source signing practices across the platform.

Although our dataset includes projects of all sizes, which yields a smaller overall commit volume than some prior studies, and does not capture repository-specific governance policies such as branch protection rules or signing requirements, these choices directly serve our broader research objective. Including very small repositories provides a more representative cross-section of the real-world developer ecosystem rather than skewing toward high-volume outliers, while abstracting away individual repository configurations allows us to reliably characterize macro-level ecosystem patterns.

Finally, while we treat commits made under different usernames by the same individual as distinct entities due to the absence of canonical identity mapping in GitHub’s API, this constraint primarily affects granular, individual-level attribution. At the aggregate scale of our study, broad ecosystem-wide trends remain dependable.

%% file: sections/8.discussion.tex
\section{Discussion and conclusion}
\label{sec:discussion}

We discuss our key management findings, compare our broader results with prior repository based studies, highlighting both common trends and oversights in past work, and offer actionable recommendations.

\subsection{Key management (RQ1)} 

\PP{Historical key tracking.} Effective key management requires maintaining a verifiable link between a developer and their keys over time~\cite{barker2018recommendation}, yet current GitHub design makes this lifecycle hard to track. Across the author accounts we analyzed, nearly one in five signed commits lacked an exact fingerprint match in the collected GitHub key registry. While push-time verification records distinguish keys that were never uploaded from previously verified keys that were later deleted, a static snapshot still cannot explain why a key was removed—such as routine rotation versus a suspected compromise. This highlights a limitation in GitHub's security model. Because users can only leave a key active or delete it without preserving revocation states or timestamps~\cite{github_commit_signature_verification}, maintainers cannot tell legitimate commits made before a breach from malicious commits signed afterward with a stolen key. Even though GitHub keeps its original verification badge after a key expires or is revoked~\cite{github_commit_signature_verification}, the reason for removal still matters. In OpenPGP, a suspected compromise casts doubt on past signatures, whereas a routine retirement does not~\cite{rfc9580}.

\begin{takeawaybox} \textbf{Recommendation.} Building on Holtgrave et al.'s proposal for signing key transparency~\cite{holtgrave2025attributing}, platforms should store historical key records, including account links, upload and deletion dates, and revocation reasons, so maintainers can apply their own trust policies during an audit. \end{takeawaybox}

\PP{Multi key management.} Using multiple keys across repositories also complicates key rotation. Many accounts use multiple key identifiers, and a substantial share return to an earlier key after adopting a newer one, often in patterns visible only across an account's combined history rather than within a single repository. 

\begin{takeawaybox} \textbf{Recommendation.} Because the latest key is not always the only active one, signing tools should support multiple authorized keys and track key usage account wide, showing where and when each key was last used. This helps maintainers check which projects still rely on an old key and verify that its replacement works before retiring it. In rare cases where one key signs commits across multiple accounts, maintainers should verify those account links and any automated workflows before revoking trust. \end{takeawaybox}

\PP{Proactive commit history auditing.} Many of the signing gaps and inconsistencies we observed likely stem from simple, everyday oversights. When developers switch machines, reconfigure local environments, or work across multiple repositories, we believe it is easy to overlook that commit signing has been disabled or that a different key is being used. While most accounts that resume signing after an unsigned gap return to their exact same key, a smaller share unexpectedly switch to a different or newly observed key. Rather than treating each commit in isolation, commit tools should audit a contributor's existing commit history to catch these mistakes before they are pushed.

\begin{takeawaybox} \textbf{Recommendation.} Commit tools (such as Git clients, IDEs, and hosting platforms) should analyze existing commit history and proactively warn users when a new commit deviates from established patterns. For example, if a user attempts to make an unsigned commit in a repository where most or all prior commits are signed, the tool should notify them. Similarly, if a user signs a commit with a key they have never used before, the tool should prompt them to confirm that the change is intentional. \end{takeawaybox}

\PP{Auditing weak cryptography at commit time.} Finally, our weak key findings show how to prioritize security upgrades. A list of uploaded keys consists largely of authentication credentials rather than commit signing keys, making static inventories a poor guide to actual signing risk. Checking key size alone also misses deprecated algorithms, as most uploaded DSA keys meet modern bit length thresholds despite being deprecated. Moreover, weak sized keys, SHA-1, and DSA signatures continue to appear in recent commits because nothing prevents users from pushing commits signed with outdated cryptography.

\begin{takeawaybox} \textbf{Recommendation.} Commit tools and hosting platforms should audit cryptographic parameters at the moment a signature is created or pushed, flagging weak key sizes, deprecated algorithms, and weak hashes. \end{takeawaybox}

\subsection{Generalizability (RQ2)}

\PP{Common trends: Low adoption.} Comparing our results with prior repository centric studies (Table~\ref{tab:rq2-comparison} in Appendix~\ref{others}) confirms that one major finding generalizes: low E2E developer signing is a platform wide challenge rather than an artifact of specific projects. Because GitHub's automatic web signing produces the majority of signed commits, raw signature counts give a false sense of security under our threat model (\S\ref{sec:threat}), masking how rarely individual developers sign their own work.

\begin{takeawaybox} \textbf{Recommendation.} To address persistently low E2E adoption, the software engineering community and hosting platforms need to raise awareness of the security guarantees of E2E signing and lower the barrier to initial adoption through clearer onboarding and education. \end{takeawaybox}

\PP{Inconsistent GitHub web signing.} Beyond confirming low overall adoption, our effort to test generalizability uncovered significant oversights in prior work. In contrast to prior studies, which assumed that GitHub automatically and universally signs all web based commits, looking more broadly revealed that platform signing is not consistent. Our audit in Table~\ref{tab:webflow_audit} shows that GitHub's web flow continued to leave some commits unsigned long after automatic signing launched~\cite{afzali2020towards}, directly contradicting documentation stating that GitHub \emph{``will automatically use GPG to sign commits you make using the web interface''}~\cite{github_commit_signature_verification}. We will disclose these findings to GitHub.

\begin{takeawaybox} \textbf{Recommendation.} GitHub should audit its automatic web signing pipeline to fix silent signing failures, or clearly document the specific conditions under which web commits remain unsigned. \end{takeawaybox}

\PP{Account level variability.} A second oversight of repository only studies is that focusing exclusively on popular repositories makes developer behavior look far more homogeneous than it is in practice. In large projects, signers appear highly consistent~\cite{holtgrave2025attributing} largely because repository rules require it~\cite{zhang2025s}. In contrast, when we follow developers across their entire accounts (\S\ref{subsection:signing_after_adoption}), we find much greater variability: most developers sign in some projects but not others, and those who sign every commit tend to have very small commit histories (Figure~\ref{fig:consistency}). Furthermore, prior large repository studies understated the adoption of newer signing technologies, finding around 1\% SSH signature adoption~\cite{holtgrave2025attributing}, whereas our account level sample (\S\ref{subsec:signing-key-results}, Table~\ref{tab:rq2-comparison}) reveals a much higher share (roughly 13\%). Developers are adopting newer, easier signing methods in smaller or personal projects long before large, established repositories reflect the shift. Tracking accounts over time also shows that when developers stop signing, they have rarely quit for good; while unsigned streaks are common, most developers later return to signing, usually with the same key.

\begin{takeawaybox} \textbf{Recommendation.} To reduce selective signing and temporary lapses across projects, Git clients and IDEs should streamline setup through built in SSH key workflows and default to global signing configurations (\texttt{commit.gpgsign = true}) once a key is registered, rather than relying on per repository setup. \end{takeawaybox}

%% file: sections/ethics.tex
\section{Ethics}
\label{sec:ethics}

\PP{Research procedures.}
We conducted our study in accordance with ethical guidelines for security measurement research and GitHub's Terms of Service and Acceptable Use Policies~\cite{github_aup}. Our analysis is purely observational: we collected only publicly available data from public repositories and user profiles via the official GitHub API, without attempting to access private repositories or non public data. All API requests complied with standard rate limits without circumvention and avoided disruptive behavior. In accordance with GitHub's information usage and privacy restrictions, we will make the resulting publication openly accessible. We did not contact users, attempt to access accounts, or modify repositories.

\PP{Stakeholders and potential harms.}
The primary stakeholders in our study are the GitHub account holders whose public profiles, commit metadata, and keys appear in our dataset, alongside the maintainers and organizations behind the analyzed repositories. GitHub operates the APIs used for data collection and provides the signature verification statuses central to our analysis, while our research team bears responsibility for safeguarding the resulting dataset. Even though every record we collected is publicly visible in isolation, aggregating and linking these records at scale introduces privacy and abuse risks by lowering the cost of profiling or targeting specific developers and projects~\cite{zimmer2020but,zook2017ten,ohm2009broken,narayanan2008robust}. Additionally, our findings carry dual use risks: highlighting repositories with signing gaps or legacy cryptographic configurations could help attackers identify vulnerable targets.

\PP{Mitigations and data stewardship.}
To balance open science and reproducibility with the protection of users, maintainers, and organizations from automated targeting, we store all collected records on a restricted access institutional server as discussed in Appendix~\ref{sec:open_science}. In our reporting, we present aggregate trends and describe the classes of signing issues we observed and why they matter for security, while omitting account names, repository identifiers, profile text, individual signing histories, and any actionable details that could be used to exploit a specific weakness.

\PP{Beneficence and justice.}
Following the principles of the Menlo Report~\cite{kenneally2012menlo}, we designed our sampling and reporting to ensure that the risks and benefits of this research are distributed fairly. We sampled accounts using IDs ordered by account creation and included repositories across a wide range of sizes so that our findings reflect everyday practices in smaller projects as well as major ones. While the developers and maintainers in our dataset bear the residual privacy risks of record linkage, they, alongside tool builders and platform designers, also stand to benefit directly from empirical evidence on how commit signing performs in practice.

\PP{Decision to publish.}
We decided to publish our aggregate findings because we believe the public interest benefits to open source software supply chain security outweigh the mitigated risks. Understanding how developers manage signing keys in day to day workflows gives tool builders concrete guidance for making key usage safer and easier. Our findings also give future interview studies a solid starting point for exploring how to make key management usable, a challenge that has plagued the security community for nearly three decades.

%% file: sections/appendix_result.tex
\section{Supplementary Results}
\label{others}

\begin{table}[h]
  \centering
  \caption{Signing key identifiers per author account.}
  \label{tab:signing-key-counts}
  \small
  \begin{tabular}{lrr}
  \toprule
  Identifiers & Accounts & (\%) \\
  \midrule
  1     & 1,896 & 60.17 \\
  2     & 626   & 19.87 \\
  3     & 286   & 9.08  \\
  4     & 149   & 4.73  \\
  5     & 80    & 2.54  \\
  6     & 53    & 1.68  \\
  7     & 19    & 0.60  \\
  8     & 11    & 0.35  \\
  9     & 7     & 0.22  \\
  $10+$ & 24    & 0.76  \\
  \midrule
  Total & 3,151 & 100.00 \\
  \bottomrule
  \end{tabular}
\end{table}

\begin{table}[h]
\centering
%\caption{%Year of the latest observed signed commit for the 12 accounts whose latest signing activity matched a weak key. Percentages use these 12 accounts as the denominator.}
\caption{Latest signed commit year for accounts last signing with a weak key ($n = 12$).}
\label{tab:latest-weak-key-years}
\small
\begin{tabular}{lrr}
\toprule
Year & Accounts & (\%) \\
\midrule
2018 & 1  & 8.33   \\
2020 & 1  & 8.33   \\
2024 & 2  & 16.67  \\
2025 & 8  & 66.67  \\
\midrule
Total & 12 & 100.00 \\
\bottomrule
\end{tabular}
\end{table}

\begin{table}[h]
\centering
\caption{Yearly signed commit counts by issue category.}
\label{tab:signing-issues-by-year}
\small
\begin{tabular}{lrrr}
\toprule
Year & Weak key size & DSA & SHA-1 \\
\midrule
2012 & 0   & 0     & 64    \\
2013 & 0   & 0     & 484   \\
2014 & 0   & 0     & 107   \\
2015 & 0   & 0     & 277   \\
2016 & 0   & 221   & 4,060 \\
2017 & 124 & 265   & 3,451 \\
2018 & 241 & 286   & 3,203 \\
2019 & 114 & 137   & 1,186 \\
2020 & 260 & 431   & 460   \\
2021 & 362 & 448   & 617   \\
2022 & 64  & 316   & 269   \\
2023 & 108 & 669   & 389   \\
2024 & 62  & 1,218 & 183   \\
2025 & 218 & 963   & 116   \\
2026 & 0   & 134   & 0     \\
\midrule
Total & 1,553 & 5,088 & 14,866 \\
\bottomrule
\end{tabular}
\end{table}

\begin{table*}
\centering
\caption{Audit of unsigned GitHub web-flow-attributed commits ($N=119,433$).}
\label{tab:webflow_audit}
\footnotesize
\begin{tabular}{lrrrr}
\toprule

\multicolumn{5}{l}{\textbf{Part A: Annual Web-Flow Signing Statistics}} \\
\textbf{Year} &
\textbf{Total Commits} &
\textbf{Signed} &
\textbf{Unsigned} &
\textbf{Unsigned (\%)} \\
\midrule
2016 & 29,826 & 0 & 29,826 & 100.00 \\
2017 & 78,674 & 17,630 & 61,044 & 77.59 \\
2018 & 104,875 & 103,243 & 1,632 & 1.56 \\
2019 & 131,245 & 128,469 & 2,776 & 2.12 \\
2020 & 190,406 & 188,061 & 2,345 & 1.23 \\
2021 & 247,811 & 244,580 & 3,231 & 1.30 \\
2022 & 338,720 & 334,913 & 3,807 & 1.12 \\
2023 & 451,989 & 448,836 & 3,153 & 0.70 \\
2024 & 707,841 & 703,658 & 4,183 & 0.59 \\
2025 & 953,261 & 946,757 & 6,504 & 0.68 \\
2026 & 88,753 & 87,822 & 931 & 1.05 \\
2034\textsuperscript{$\dagger$} & 1 & 0 & 1 & 100.00 \\
\midrule
\textbf{Total}
& \textbf{3,323,402}
& \textbf{3,203,969}
& \textbf{119,433}
& \textbf{3.59} \\

\midrule
\multicolumn{5}{l}{\textbf{Part B: Temporal Distribution of Unsigned Web-Flow Commits}} \\
\textbf{Period} &
\multicolumn{2}{r}{\textbf{Unsigned Commits}} &
\textbf{Unsigned Rate (\%)} &
\textbf{Share (\%)} \\
\midrule
2016--2017 &
\multicolumn{2}{r}{90,870} & 83.75 & 76.08 \\
2018--2023 &
\multicolumn{2}{r}{16,944} & 1.16 & 14.19 \\
2024--2026\textsuperscript{$\dagger$} &
\multicolumn{2}{r}{11,619} & 0.66 & 9.73 \\

\midrule
\multicolumn{5}{l}{\textbf{Part C: Verification and Attribution Audit ($N=119,433$)}} \\
\textbf{Characteristic} &
\multicolumn{2}{r}{\textbf{Count}} &
\multicolumn{2}{r}{\textbf{Proportion (\%)}} \\
\midrule
\multicolumn{5}{l}{\textit{Verification metadata}} \\
Reason: \texttt{unsigned} &
\multicolumn{2}{r}{119,433} &
\multicolumn{2}{r}{100.00} \\
Verified: false &
\multicolumn{2}{r}{119,433} &
\multicolumn{2}{r}{100.00} \\
\addlinespace
\multicolumn{5}{l}{\textit{Committer attribution}} \\
Committer name: \texttt{GitHub} &
\multicolumn{2}{r}{119,433} &
\multicolumn{2}{r}{100.00} \\
Missing author login &
\multicolumn{2}{r}{3} &
\multicolumn{2}{r}{0.003} \\

\bottomrule
\end{tabular}

\vspace{2mm}
\begin{minipage}{0.95\textwidth}
\footnotesize
\textit{Notes:}
All commits are classified as Web (\texttt{web-flow}) based on \texttt{committer\_login = web-flow} and \texttt{committer\_name = GitHub} (6 additional unsigned commits with \texttt{committer\_login = web-flow} but non-\texttt{GitHub} committer names are classified under \textit{Other} in Table~\ref{tab:signing_breakdown}). Signing status is determined by the presence of a signature. In Part B, unsigned rate denotes the proportion of web-flow commits lacking signatures within each period, whereas share denotes the proportion of all 119,433 unsigned web-flow commits. \textsuperscript{$\dagger$}Includes one commit with a future-dated timestamp (2034-12-29); in Part B, this commit is grouped with the 2024--2026 period (11,618 from 2024--2026 plus 1 from 2034 = 11,619).
\end{minipage}
\end{table*}

\begin{table*}
\centering
\caption{Descriptive comparison of prior findings with account-sampled observations
of public GitHub activity.}
\label{tab:rq2-comparison}
\footnotesize
\setlength{\tabcolsep}{4pt}
\renewcommand{\arraystretch}{1.0}
\begin{tabularx}{\textwidth}{@{}
  >{\raggedright\arraybackslash}p{.13\textwidth}
  >{\raggedright\arraybackslash}X
  >{\raggedright\arraybackslash}X@{}}
\toprule
\textbf{Measure} & \textbf{Prior findings~\cite{holtgrave2025attributing,sharma2025commit,zhang2025s}} & \textbf{Our findings} \\
\midrule
\textbf{Signing prevalence}
& Holtgrave et al.\ found 15.7\% signed overall
  and 5.0\% outside \texttt{web-flow}. \texttt{web-flow}
  supplied 72.89\% (2,961,704/4,063,376) of signed commits.
& 24.05\% (3,874,732/16,112,439) signed overall, with 82.69\%
  (3,203,969/3,874,732) of signed commits originating from
  \texttt{web-flow} (Tables~3 and~\ref{tab:signing_breakdown}).
  5.22\% (658,217/12,617,006) signed in the restricted E2E cohort.
  Both samples show low signing outside \texttt{web-flow}. \\
\addlinespace[2pt]

\textbf{Adoption and consistency}
& Holtgrave et al.\ grouped committers by email:
  95.4\% never signed and 2.0\% signed every commit. Among
  signers, Fig.~12 (right) shows a median near 85\%.
& 5.15\% (3,151/61,154) of E2E author accounts had a signed commit.
  0.44\% (270/61,154) had all observed E2E commits signed
  (8.57\% of signers). The median signing rate among signers was 29.13\%. \\
\addlinespace[2pt]

\textbf{Signature formats}
& Holtgrave et al.\ found 98.87\% PGP, 1.08\% SSH, and
  0.05\% S/MIME among 1,101,672 non-\texttt{web-flow} signatures.
& Of 658,217 restricted E2E signatures, 86.60\% (569,985) were GPG,
  13.11\% (86,287) SSH, and 0.30\% (1,945) CMS/S/MIME (Table~8). The formats rank in
  the same order, with a substantially higher SSH share in our sample. \\
\addlinespace[2pt]

\textbf{Repository variation}
& Holtgrave et al.\ found 72.1\% of repositories
  lacked a non-\texttt{web-flow} signature and observed lower signing
  fractions in more active repositories. Sharma et al.\ (Secs.~1 and 5.1)
  reported 2.55\% verified in machine learning, 5.95\% in databases,
  28.35\% in security, and 12.77\% in web development.
& Signing fractions varied across repositories for 2,841 of 2,990
  multi-repository signers (95.02\%). Median signing was 0\% in
  both activity bins over 79,269 account--repository pairs.
  These are within-account comparisons. \\
\addlinespace[2pt]

\textbf{Timing and resumption}
& Sharma et al.\ observed an initial rise in verified shares
  after April 2021 and later declines in several domains. In
  Zhang et al., 3 of 5 prior signers said they would continue only
  if projects required it, citing key-management effort.
& 453 of 3,131 eligible signers (14.47\%) signed their first observed E2E commit.
  Late adopters had a median of 132 earlier unsigned commits.
  34.02\% (1,072/3,151) had an unsigned tail after their last
  observed signature. Within 365 days of the first observed
  signature, 1,063 of 1,448 (73.41\%) signed again after an unsigned
  commit. The adjusted 365-day odds ratio per account year was
  1.011 (95\% CI: 0.987--1.037; $p=0.3731$). \\
\addlinespace[2pt]

\textbf{Account attribution}
& Holtgrave et al.\ found 30.1\% author, committer, and
  pusher agreement in its push-event subset. Zhang et al.\ found merges can resemble impersonation. Sharma et al.\ observed attribution through unverified email.
& We require author--committer account agreement (\texttt{author\_login = committer\_login}) and exclude \texttt{web-flow} for restricted E2E analyses. \\
\addlinespace[2pt]

\textbf{Registered and observed keys}
& Holtgrave et al.\ found 22.7\% of linked accounts
  had uploaded PGP or SSH signing keys, typically one or two.
& 1,896 of 3,151 signers (60.17\%) had one observed key identifier (Table~10).
  Full fingerprints matched the collected registry for
  540,726 of 658,217 signed commits (82.15\%; Table~8). \\
\addlinespace[2pt]

\textbf{Keys across accounts}
& Holtgrave et al.\ found 25 instances of keys shared between
  GitHub accounts in its uploaded-key analysis.
& 3 of 6,115 distinct format--fingerprint pairs (0.05\%) occurred
  on signed commits attributed to multiple accounts (spanning 8 accounts). \\
\addlinespace[2pt]

\textbf{Key changes and reuse}
& The three studies provide no corresponding fingerprint-sequence
  analysis across repositories or signing gaps.
& Of 1,255 accounts with multiple identifiers, 679 (54.10\%) changed keys
  with no observed return, 414 (32.99\%) returned to an earlier key, and 162 (12.91\%) were unresolved.
  Of 1,047 eligible multi-repository accounts, 975 (93.12\%) reused keys across repositories
  and 758 (72.40\%) used multiple keys within a repository. Furthermore, 781 of 1,063
  accounts (73.47\%) resumed signing with the same fingerprint after their first
  observed run of unsigned commits. \\
\addlinespace[2pt]

\textbf{Legacy GPG signing}
& Holtgrave et al.\ found 626 uploaded DSA-1024 PGP keys.
& Of 569,985 GPG-signed commits, 5,088 (0.89\%) used DSA (latest
  Mar.\ 11, 2026) and 14,866 (2.61\%) used SHA-1 (latest Dec.\ 4, 2025).
  We also matched 1,553 GPG signatures (0.24\% of all signed commits) to weak-size ($<$2,048-bit RSA-equivalent) uploaded key
  components (Tables~9 and~12). These counts can overlap. \\
\bottomrule
\end{tabularx}

\smallskip

\begin{minipage}[b]{\textwidth}
\raggedright
Holtgrave et al.~\cite{holtgrave2025attributing} studied 50,328
dependency-selected projects (25.89M commits).
Sharma et al.~\cite{sharma2025commit} studied 60 popular repositories
(869,437 commits).
Zhang et al.~\cite{zhang2025s} sampled 9,810 repositories via
star-ranked searches (12.58M commits).
\end{minipage}\par
\end{table*}